\documentclass{anabs}
\usepackage{graphicx}
\usepackage{psfig}
\usepackage{times}
\Volume{1-2}              
\Year{2003}              
\Month{02}               
\Pagespan{000}{000}      

\begin{document}
\lhead[\thepage]{A.N. Author: Title}
\rhead[Astron. Nachr./AN~{\bf 324} (2003) 1/2]{\thepage}
\headnote{Astron. Nachr./AN {\bf 324} (2003) 1/2, 000--000}

\title{The ASCA Slew Survey}

\author{Ken Ebisawa\inst{1,2}, 
R. Fujimoto and Y. Ueda\inst{3}}
\institute{
INTEGRAL Science Data Center, Chemin d'\'Ecogia 16, CH-1290, Versoix, Switzerland
\and 
Laboratory for High Energy Astrophysics, NASA/GSFC, Greenbelt, MD 20771, USA
\and 
Institute of Space and Astronautical Science, Yoshinodai, Sagamihara, Kanagawa 229-8510, Japan}

\correspondence{ebisawa@obs.unige.ch}

\maketitle

\section{Introduction}
We are  systematically analyzing ASCA GIS data taken during the
satellite attitude maneuver operation. 
Our motivation is to search for serendipitous hard X-ray sources 
and make the ASCA Slew Survey catalog.  

 Although ASCA is not designed to carry out observations
during  attitude maneuver/slew operation, 
the GIS instrument was operational  and  collecting X-ray events
during most attitude maneuvers.  GIS has a high time resolution
($\sim$ 60 msec)
so that it can determine the photon arrival direction while the 
satellite is moving (maximum maneuver velocity is  $\sim$ 0.2 deg/sec).  
 During its operational life from 1993 February to 2000 July, ASCA
carried out more than 2,500 maneuver operations, and total exposure time
during the maneuver was $\sim$ 415 ksec after proper GIS data screening.

In total, $\sim$ 60 \% of the sky (7.6 str) was scanned at least once,
and the total exposure time $\times$ area is $\sim$ 60 str $\cdot$ sec.
Therefore, average exposure time of the scanned region is  $\sim $ 7.9 sec.
See  Figure 1 for the exposure map.  

\begin{figure}
\centerline{\psfig{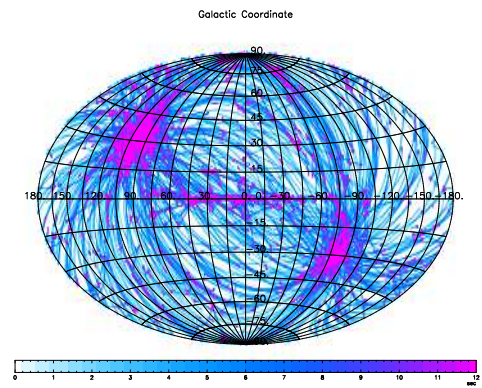}}
\caption{
Exposure map of the ASCA GIS slew data in Galactic coordinates.
Different color indicate the exposure time for each bin.
Note that most scans are along the ecliptic meridian due to the
ASCA operational constraint.  Consequently, we have good coverage
in the north and south ecliptic regions.}
\label{label1}
\end{figure}

\section{Preliminary Results}

\begin{figure}
\centerline{\psfig{file=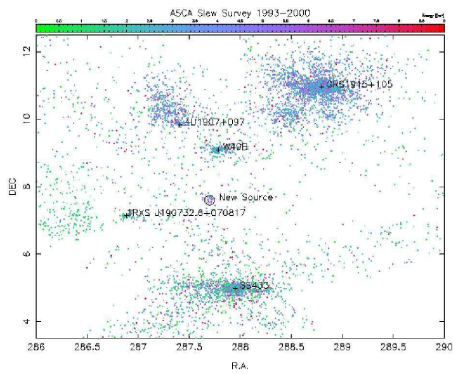,height=5.5cm,angle=-90}}\caption{
Close-up of the slew data in the GRS1915-105 -- SS433 region. 
 A new hard source is discovered at
$(\alpha,\delta)$ =(287.6, 7.7)  in the north of  SS433.}
\end{figure}

We found the most technically challenging  part in this project
is to determine the satellite attitude  during the attitude maneuver
accurately,
since this is beyond the 
specification of the  ASCA attitude control and determination system which is designed 
only for the pointing observations.  We have been working with the engineers who have designed and built
the ASCA attitude control and determination system.  

Using the thus reconstructed  satellite attitudes, we have processed all the maneuver data,
and determined positions of GIS events.
Average exposure time is $\sim$ 8 sec/FOV for a scanned region, during which
 we may collect $\sim$ 10 events for a  0.625 cts/GIS source  using two GIS;
this  corresponds to $\sim$ 0.6 mCrab, and considered to be 3 $\sigma$ sensitivity of our dataset.

We expect to discover new hard X-ray sources
which may be bright only above 2 keV and have not been detected by the ROSAT all sky survey.
Actually, we   did discover a new hard X-ray source at $(\alpha,\delta)$
=(287.6, 7.7)  in the north of  SS433.
Find the contrast in color (hardness) with the nearby ROSAT all sky survey source 
(1RXS J190732.6+070817).

\end{document}